\UseRawInputEncoding
\documentclass[10pt,journal,cspaper,compsoc]{IEEEtran}

\usepackage{cite,url}
\usepackage{graphicx}
\usepackage{epstopdf}
\usepackage{amssymb,amsfonts,stmaryrd}
\usepackage{pifont}

\begin{document}

\title{Service Composition in Opportunistic Networks: A Load and Mobility Aware Solution}

\author{Umair~Sadiq, Mohan~Kumar, Andrea~Passarella and Marco~Conti
\IEEEcompsocitemizethanks{\IEEEcompsocthanksitem U.~Sadiq is with the Sabre, USA.\protect\\ E-mail: umair.sadiq@sabre.com. 
\IEEEcompsocthanksitem M.~Kumar is with the Rochester Institute of Technology, USA.\protect\\ E-mail: mjkvcs@rit.edu

\IEEEcompsocthanksitem A.~Passarella and M.~Conti are with the IIT-CNR, Pisa, Italy.\protect\\
E-mail: \{a.passarella, m.conti\}@iit.cnr.it
}
\thanks{}
}

\markboth{IEEE TRANSACTIONS ON COMPUTERS,~Vol.~63, No.~9, September~2014}%
{Sadiq \MakeLowercase{\textit{et al.}}: Service Composition}

\IEEEcompsoctitleabstractindextext{%
\begin{abstract}

Pervasive networks formed by users' mobile devices have the potential to exploit a rich set of distributed service components that can be composed to provide each user with a multitude of application level services. However, in many challenging scenarios, opportunistic networking techniques are required to enable communication as devices suffer from intermittent connectivity, disconnections and partitions. This poses novel challenges to service composition techniques. While several works have discussed middleware and architectures for service composition in well-connected wired networks and in stable MANET environments, the underlying mechanism for selecting and forwarding service requests in the significantly challenging networking environment of opportunistic networks has not been entirely addressed. The problem comprises three stages: i) selecting an appropriate service sequence set out of available services to obtain the required application level service; ii) routing results of a previous stage in the composition to the next one through a multi-hop opportunistic path; and iii) routing final service outcomes back to the requester. The proposed algorithm derives efficiency and effectiveness by taking into account the estimated load at service providers and expected time to opportunistically route information between devices. Based on this information the algorithm estimates the best composition to obtain a required service. It is shown that using only local knowledge collected in a distributed manner, performance close to a real-time centralized system can be achieved. 
Applicability and performance guarantee of the service composition algorithm in a range of mobility characteristics are established through extensive simulations on real/synthetic traces.
\end{abstract}

}

\maketitle
\IEEEdisplaynotcompsoctitleabstractindextext
\IEEEpeerreviewmaketitle

\ifCLASSOPTIONcompsoc
  \noindent\raisebox{2\baselineskip}[0pt][0pt]
  {\parbox{\columnwidth}{\section{Introduction}
  \global\everypar=\everypar}}
  \vspace{-1\baselineskip}\vspace{-\parskip}\par
\else
  \section{Introduction}\par
\fi

\IEEEPARstart{I}{n} recent years, the number of multi-functional, personal smart devices has been increasing at a high rate. The possibility of such devices coming within communication range of each other is enhanced by the presence of multiple embedded radios. While such opportunistic contacts between pairs of devices have been exploited by the opportunistic networking paradigm \cite{pelusi,spyroscp:17,ebr:14}, exploiting resources to execute remote services is an area of research that has received attention in recent times \cite{Passarella,servexec}. Heterogeneous resources available on devices can be abstracted as services to simplify the interface and have platform independence. Services from multiple devices can be composed to provide enhanced functionality \cite{svccomp,opportunities}. 
For example, consider the following scenario: an event takes place at some location and we need to collect and process all related data available on devices carried by people in the physical surroundings. Using the devices in physical proximity of the event, we establish an event context, use this context to filter relevant data, perform necessary processing on filtered data and then finally collect processed data such that each of these steps can be performed on separate devices in sequence depending on available resources.  Each task in the above scenario is a service that is executable on a device. For example, a jogger in a park wishes to avoid dogs and areas with high 
 density of pollens:  her device requests information from other joggers 
 (initiating service); user devices respond (services) by filtering the 
 raw data available on their devices, possibly composing data received 
 from other devices nearby (this may entail many service components, such 
 as transmit pictures, live streams, tweets,
 filtered information on dog barking, bees etc.). More in general, 
 service composition and opportunistic invocation is applicable in pervasive healthcare, intelligent transportation systems, crisis management, etc.~\cite{opportunities}, and environments like parks, malls, streets in a city or other social gathering places where citizens interact. In general, service composition can be applied whenever a mobile device needs a hardware or software resource that is not available locally, but that can be found in its proximity, on other devices around that can be reached through opportunistic networking techniques. Nowadays, this type of collaborative use of resources is not possible, as mobile devices either need to have all functionalities locally, or have to rely on infrastructure-based communications to reach services over the Internet. The availability of local services on other devices is not exploited.

Service composition in opportunistic networks is critical, e.g., consider following scenarios: (i) required services may not be available over the Internet -- local location coordinates for a device without GPS, (ii) downloading content from Internet consumes (limited, costly) bandwidth -- a nearby device may provide a similar service through freely available open spectrum, or (iii) communication to the Internet itself may not be available -- remote location, bad signal or crisis management scenarios etc. The creation of a services-rich environment by effectively making services 
available on each device, accessible to applications on other devices in opportunistic environments  is a major challenge. The novelty of our solution is that service composition is performed without any requirements for a connected path between service providers and the service requester (see Section~\ref{relatew} for a discussion on related work). In particular, an efficient mechanism for selecting services to complete a request and forward service parameters to corresponding devices is proposed.

Our algorithm selects a service sequence set by taking into account the service load and  temporal distances between nodes (temporal distance provides a measure of relative location of other nodes). These values are measured in  a distributed manner by using only opportunistic contacts. The elegance of our solution lies in the fact that while a particular service sequence is selected by considering the temporal distance between devices, the actual routing scheme used to forward service request and parameters can be different. For example, to meet a service request of \textit{encrypting and compressing} a file, our algorithm chooses a sequence of services (e.g., an encryption service and then a compression service) that can be composed such that these services are provided at devices in proximity of the service requester. \footnote{Specifically, devices whose distance from the requester (a proxy for the time to reach them thorough a multi-hop path) and load is such that the service can be provided in a short amount of time. Our simulation results shows that our heuristic is effective in finding the best solution.} Our proposed service composition uses an underlying routing scheme to forward service request to the device providing encryption, encrypted result to the device providing compression and lastly to transmit final result back to the service requester. 

Extensive simulation results are presented to demonstrate the performance of the proposed method in terms of service composition success, composition length, number of hops, and delays. The performance is analyzed for a range of service densities, number of nodes, service request rates, request timeout durations, and  routing mechanisms. It is shown that: i) composition is better than searching for the exact single service match; ii) use of multi-hop paths improves performance (service completion rate and delay) significantly as compared to one-hop direct forwarding; and iii) using only local knowledge (collected from opportunistic contacts) about load at and temporal distance from other nodes, performance close to a centralized system (exact load and temporal distances between nodes are known) can be achieved. This paper is an extension of our prior work\cite{ourwork}. Key additional contributions are as follows. Scope and applicability of our service composition algorithm is established by considering a range of mobility characteristics - movement in communities, clustered environments, uniformly distributed mobile users [Section \ref{mobCharac}], analyzing  relationship between estimated cost in the service graph and actual delay to complete the service composition request [Section \ref{perfGuarant}], and evaluating sensitivity to composition length and service distribution [Section \ref{sensitivity}].

In the rest of the paper, the terms {\em devices} and {\em nodes} are used interchangeably. The paper is organized as follows: Section~\ref{relatew} discusses related work. Section~\ref{xsysdesc} provides the system description. Service composition mechanism and algorithm is described in Section~\ref{sercomp} followed by modeling of temporal distance and service loads in Section~\ref{resModeling}. Performance of service composition algorithm is evaluated in Section~\ref{perfeval}. Specific characteristics of the algorithm and sensitivity to key factors are analyzed in Section~\ref{xcharcal}  and Section~\ref{xconclusion} concludes with directions for future work.

\section{Related work}

\label{relatew}

Here we provide a description of existing research in service representation, and execution of single and multiple services.

\textbf{Service representation:} There has been a significant research on semantics and ontologies to describe services and compositions (see \cite{scmobile} for survey and issues of service composition in mobile environments). Services to be composed can either be strictly defined in the request (static composition) \cite{spidernet},  or can be computed at run-time (dynamic composition) based on the request \cite{svccomp}.  Dynamic composition looks for alternate sets of services in the current environment that can be composed to provide the required functionality.  Services are modeled as directed attributed graphs \cite{svccomp}, to find possible compositions.  We leverage this work for service representation to find possible compositions and explore mechanism to execute such compositions in an opportunistic network. 

\textbf{Requesting Single Services:} Passarella et al., \cite{Passarella,servexec} investigated optimal policy for service execution in opportunistic networks. An optimal policy is derived for the level of replication to receive service results in minimal time but only a single service is requested and the service requester waits until a direct contact with service provider. Also, a few other works propose efficient schemes and use multi-hop paths for service discovery and invocation in opportunistic networks by using a set of proxies \cite{provision1} or location of the user \cite{provision2}. However, composition of multiple services is not considered in these works. 

\textbf{Composing Multiple Services:} For composition of services, a number of middleware frameworks and architectures are proposed but these do not discuss the actual forwarding mechanism \cite{svccomp,arch}. The use of multi-hop paths to send service requests and receive service outputs between nodes is considered in \cite{reliable,dependable} by modeling the expected colocation time between nodes based on past interactions. However, it needs to be a connected path and service composition fails if participants do not remain directly connected. Moreover, in open environments, there may not be any history of past interactions which can be used to predict which devices will be connected at what time in future. 

In contrast to existing solutions, in this paper, service composition is performed using multi-hop paths that can relay service inputs to the service providers and then relay service results back to the service requestor {\em even when an end-to-end connected path does not exist}. None of the existing solutions enable service composition using opportunistic paths. Furthermore, alternative paths are considered (based on the service load on devices and the temporal distances between devices) to increase the success of composition.

\begin{figure*}[!t]
\centering
\includegraphics{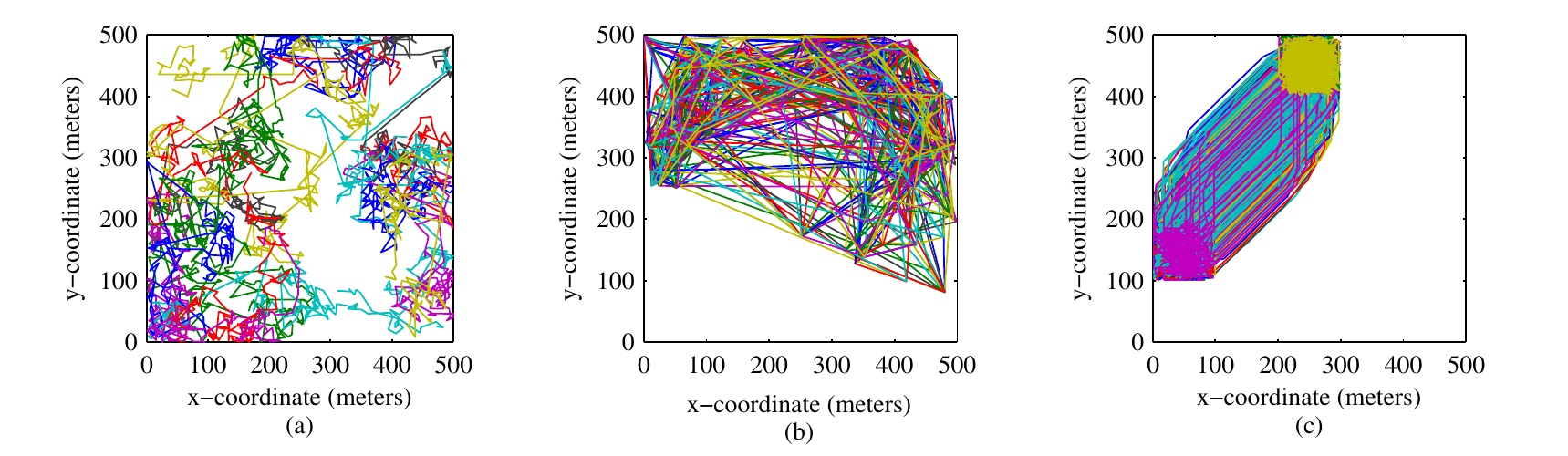}   
\caption{Trajectories of 20 users in (a) Levy walk, (b) SLAW, and (c) HCMM mobility models.}
\label{1_trajectories}
\end{figure*}

\section{System description}
\label{xsysdesc}
\subsection{Service model}
\label{serviceModel}

We employ a service graph 
to represent services, their inputs and outputs, and concatenation of services
 based on their input requirements. This can be done by using service ontologies and semantics in a hierarchical manner \cite{svccomp}. For simplicity, we represent a service based on its input and output types to generate a service graph i.e.\ two services can be composed if output type of one is same as required input to the other. For $n_d$ input/output types, possible services are of type $s_{xy}$ such that input type $x\in[1,n_d-1]$ and output type $y\in[x+1,n_d]$. This representation illustrates services with different levels of functionality. A service $s_{xy}$ is of functionality $k$ if $y=x+k$. Suppose, $s_{12}$  and $s_{23}$ represent services \textit{decompress} and \textit{decrypt} respectively, then service $s_{13}$ represents a service of higher functionality that can perform both decryption and decompression. A service with a higher functionality such as $s_{14} $ provides same transformation that is otherwise possible by composition of two or more services with lower functionalities (e.g., $\{s_{12}, s_{23}, s_{34}\},\{s_{12}, s_{24}\}$ or $\{s_{13}, s_{34}\}$). Services with varying functionalities are used to demonstrate the tradeoff between selecting an exact service (that can be further away in the network) versus composition of smaller services (that might be available in proximity of service requester). To account for task and device heterogeneity, we assume that execution time of each service is distributed exponentially on the devices. The performance of service composition is explored by varying the repetition of each service i.e.\ the number of nodes providing that service.

\subsection{Mobility model}
Real mobility traces are shown to have a power law distribution of Inter-Contact Times (ICTs) in contrast to previously assumed exponential distribution \cite{mobimpact:4}. Since random waypoint and random direction models lead to exponential Inter-Contact Times \cite{scalinglaws:27}, they are not used in this paper to represent nodes' mobilities. In addition, it has been shown that real mobility traces have power-law flight-lengths and pause-time distributions \cite{levywalk:28}. Therefore, in our analysis and simulation, we generate synthetic traces using Levy walk mobility model \cite{levywalk:28} and SLAW mobility model \cite{slaw:29}. Levy walk and SLAW represents a more realistic scenario in comparison with \cite{servexec} where each node is equally likely to meet any other node. In addition, we also evaluate performance using the HCMM mobility model \cite{boldrini2010hcmm} that models spatial and temporal properties of human mobility by also taking into account that user mobility is driven by existing social relationships between them.

Mobility plays a central role in the selection and forwarding of service requests in an opportunistic network. These models provide a wide range of mobility characteristics to thoroughly evaluate performance of service composition: HCMM - movement in communities; SLAW - a clustered environment; and Levy walk - uniformly distributed mobile users. These characteristics are also reflected in Figure~\ref{1_trajectories} that shows trajectories of 20 users under these models. Flight lengths have power law distribution and user mobility is spread in the entire area under Levy walk whereas only specific sites (called waypoints \cite{slaw:29}) are visited under SLAW. As a result user movement is more evenly spread in Levy walk [Fig.~\ref{1_trajectories}a] and contains clusters in SLAW [Fig.~\ref{1_trajectories}b]. Fig.\ref{1_trajectories}c shows the extreme case where users only move between two communities where this movement is controlled by the parameter of rewiring probability in HCMM \cite{boldrini2010hcmm}. 

\subsection{Forwarding schemes}
\label{formet}

There are several forwarding schemes proposed in the literature on opportunistic networks. The following forwarding schemes have been found to have good performance in dynamic environments(see \cite{proximol} for detailed discussion): i) use of timer transitivity (TT) \cite{spyroscp:17} or temporal distance between nodes in homogeneous (nodes with similar mobility characteristics) environments; and ii) use of encounter rate (EBR)\cite{ebr:14,hetnodesspyro:15} in heterogeneous environments (nodes with varied mobility characteristics). In addition, we have proposed a modified version of timers (MT) \cite{proximol} that works efficiently over small forwarding paths. In our simulation studies, we compare performance of the proposed  selection and forwarding mechanism for service composition under 4 forwarding schemes - direct or one-hop, TT, EBR, and MT. Direct forwarding belongs to the class of one-hop routing schemes whereas, TT, EBR and MT belong to the class of multihop routing schemes.

\begin{figure*}[!t]
\centering
\includegraphics{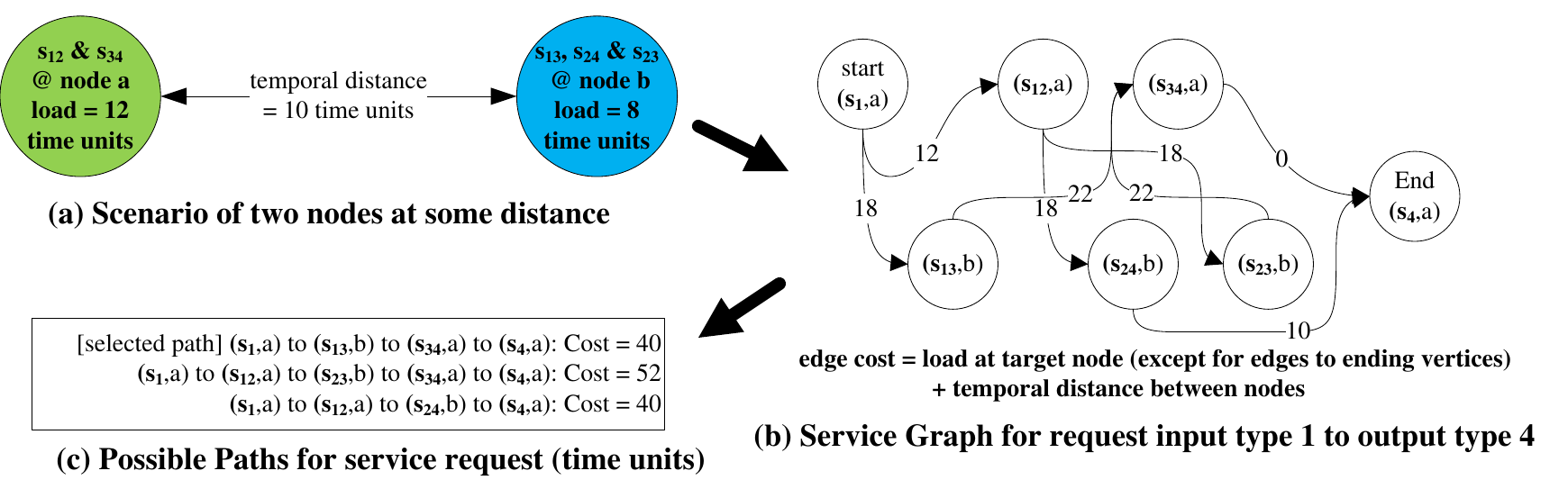}   
\caption{Overview of Composition Selection Algorithm: (a) an example scenario, (b) construction of service graph, and (c) possible paths for request completion}
\label{figsim1}
\end{figure*}

\section{Service composition algorithm}
\label{sercomp}
In this section we provide an overview of the service composition algorithm. Each device maintains a service graph and the link costs on the service graph are updated based on temporal distance between nodes and their current service loads. A particular service composition is selected based on the shortest path in this graph. Overview of the key algorithm steps is also provided in Figure \ref{figsim1}. Resource modeling for temporal distance and load as wells as methods for estimation are provided later in Section \ref{resModeling}.

\subsection{Service graph}
Each device maintains a local service graph $G=(V,E)$ based on its view of rest of the network. A simple case is described in Figure \ref{figsim1}, where services provided by two devices \textit{a} and \textit{b} are used to construct a service graph. The number of devices, services (including repetitions) and input/output types are represented by $N, N_s$ and $n_d$ respectively. The service graph $G$ has two types of vertices $V = \{V_1, V_2\}$, where a vertex ${v \in V_1}$ is a device and service pair such that $|V_1| = N_s$, and vertex ${v \in  V_2 }$ is a device and input/output type pair such that $|V_2| = n_d$. The two types of the vertices are shown in Figure \ref{figsim1}b: i) $(s_{12},a)$ is a vertex of type $V_1$, represents that service $s_{12}$ is provided at node $a$;  and ii) $(s_1,a)$ is a vertex of type $V_2$, represents that there is an input/output of type $1$ (denoted $s_1$) entering/exiting an application at node $a$. Vertices of type $V_2$ are only maintained for the device constructing the service graph.

A directed edge from vertex $u$ to vertex $v$ exists if (i)  input/output type of $u\in V_2$ is same as service input of $v\in V_1$ e.g., $(s_{1},a)\shortrightarrow (s_{12},a)$; (ii) service output of $u\in V_1$ is same as service input of $v\in V_1$ e.g., $(s_{12},a)\shortrightarrow (s_{24},b)$; or (iii) service output of $u\in V_1$ is same as input/output type of $v\in V_2$ e.g., $(s_{24},b)\shortrightarrow (s_{4},a)$. The cost of an edge  in cases (i) and (ii) is the sum of temporal distance between devices of corresponding vertices  and the load at the device of second vertex. For example, cost of $(s_{12},a)\shortrightarrow (s_{24},b)$ is $18$ which is sum of temporal distance from $a$ to $b$ $(10)$ and load on $b$ $(8)$. However, the cost of an edge in case (iii) is simply the temporal distance between devices of corresponding vertices as it is the cost of routing  final results back to the service requester.  Vertices of type $V_2$ are needed to take into account the temporal distances: (a) from the service requestor to the node providing the first service; and (b) from the node providing the last service to the requestor.

\subsection{Algorithm}
Our final algorithm is to compute a shortest path on a service graph based on a service request (input type and desired final output type). From the collected information about service loads, temporal distances, and services provided at the neighboring nodes, a device creates the service graph. In Figure \ref{figsim1}, a simple case is described  for a service request from input type $1$ to output type $4$ at device \textit{a}. Using the service load and temporal distance estimate, the edge cost is computed from each service output to compatible service inputs. From the final graph, shortest path is computed from the starting vertex $(s_{1},a)$ to ending vertex $(s_{4},a)$ by Dijkstra's Algorithm. In this case the shortest path is $(s_{1},a)\shortrightarrow (s_{13},b) \shortrightarrow (s_{34},a) \shortrightarrow (s_{4},a) $ i.e.\ the next service $s_{13}$ is to be run at device \textit{B}. This information is given to the forwarding algorithm to route service request from device \textit{A} to device \textit{B}. After execution of this service in the sequence, the device hosting the service (device \textit{B}) finds the next service in the path by re-computing the shortest path for pending request (input type $3$ to output type $4$) and gives the destination address to the underlying forwarding scheme. Service execution and re-computing the shortest path for pending request is done until the final results are routed back to the service requestor. The path is re-computed after execution of each service in the sequence as the network topology can change in that duration to provide alternate paths that are more feasible. However, one may choose to follow the same path computed at the service requestor if the network topology does not change very fast.

The network overhead in terms of data shared upon each contact between two nodes is only $O(N)$ values. These values comprise of one estimate for load and temporal distance for each device in the network. The service graph has a total of $N_s+n_{d}$ vertices. The computational complexity of updating the service graph requires updating $O(N^2)$ values (of edge cost) instead of $O((N_s+n_d)^2)$ values because the cost of edges from  services on one device to services on another is same. Also, Dijkstra's algorithm has a computational complexity of $O((N_s+n_{d})^2)$. One run of Dijkstra's Algorithm finds shortest path from one input type to all possible  output types i.e.\ it finds composition paths of all service requests that have the same input type. Thus, at most, a node only requires to run Dijkstra's Algorithm $n_{d}$ times to compute composition path for any number of pending service requests. To control network overhead and computational complexity in a large network, the size of graph can be reduced by retaining information of only those devices that are in close vicinity.


\section{Resource modeling and parameter estimation}
\label{resModeling}

\subsection{Disconnected paths}
In opportunistic networks, nodes providing a service may not be directly connected to the service requested. This leads our investigation into modeling the physical distance between nodes when these are devices carried by people. 
It is shown in \cite{proximol} that the physical distance between nodes is related to the separation in time, i.e.\ time since they moved out of each other's transmission range. Also, nodes can share information about separation in time to create a local view of the entire network (i.e.\ distance / separation in time from every other node in the network). A mechanism to find and share such values (separation in time) has been proposed in \cite{spyroscp:17} and the final computed values for separation in time is called (and is equivalent to) shortest temporal distance. Several works have  used the intuition behind this idea to approximate the physical distance between nodes and characterize the temporal distance \cite{proximol,spyroscp:17, agingrules:24, agematter:31, lastenc:32, tempordist:38}. Besides its relation to the physical distance, an alternate definition of shortest temporal distance between two disconnected nodes is that it is the minimum time in which some sequence of contacts can relay information from one node to the other \cite{tempordist:38}. This is described in the next section. 

\subsection{Shortest temporal distance}
\label{stdr}
In essence, shortest temporal distance $(t_{ij})$ provides the minimum time (using epidemic forwarding) it would take for some information to travel from node \textit{i} to node \textit{j}. In this paper, the term temporal distance always refers to the shortest temporal distance. Therefore, to compare efficiency of an underlying forwarding scheme to send requests to different nodes in network, we simply compare this lower bound on the propagation time. For example, node  \textit{i} can select a service provider node \textit{j} or \textit{k} based on which of these has smaller distance ($t_{ij}$ or $t_{ik}$). Since,  no centralized infrastructure exists in an opportunistic network to provide $t_{ij}$ and $t_{ik}$, the approximations $t_{ji}$ and $t_{ki}$ are used as they can be measured at node \textit{i} in a distributed manner by using simple timers [Section \ref{timerx}]. It is shown in \cite{proximol} that the physical distance between nodes is related to temporal distance. As the distance between nodes $i$ and $j$ is the same in both directions, the temporal distances are also approximately equal in both directions. 

\subsection{Service loads}
Each node provides one or more services, and maintains a FIFO queue for all service requests. Thus, even though a requester may be connected to a service provider, the request may take a long time to complete because service provider is busy. Therefore, one needs to take into account the current service load at the destination in order to compose services in minimal time. This can be done by sharing service load of all nodes in a distributed manner [Section \ref{timerx}]. Our service selection algorithm takes into account the service load at destination together with the temporal distance estimate. Though, statistical mechanisms can be used to predict the future service loads and provide better estimates, our current analysis uses a moving time-window average for the load estimate $l$ at each node with a decaying factor $\alpha$ of 0.5. 
Load in current window $l_{cw}$ is calculated by the total number of pending requests multiplied by the expected service execution time. Then estimate of load is updated by the following rule: $l = \alpha l_{cw} + (1-\alpha)l_{old}$ where $l_{old}$ is the load value $l$ in previous time window. We use a time window of 30s in our analysis.  

\subsection{Estimation of temporal distance and load}
\label{timerx}
We are interested in temporal distance between nodes. An estimate of this distance at some time $t_0$ can be used to approximate it for a later time that is not too long after $t_0$. In order to estimate temporal distance from other nodes in a distributed manner, each node keeps a timer for every other node in network using an update rule defined below. Nodes also share their load values with rest of the network. This is in contrast with centralized estimates of load that are used in \cite{servexec}. Each node maintains a load value for other nodes in network. Let $t_a(i)$ denote the time elapsed since node \textit{a} last made contact with node \textit{i}, where $t_a(a)$ is always set at zero. Also, let $l_a(i)$ denote the estimate for load of node \textit{i} at node \textit{a}.   Local timer values for each node are incremented after every time unit (e.g., 30s, 60s etc.\ depending on devices). When node \textit{a} comes into contact with some other node \textit{b}, it updates its timers and loads estimates for all nodes from whom node \textit{b} has a smaller temporal distance. This rule is defined as follows:  $\forall i \not = a$ if $t_b(i) < t_a(i) - t_{av}$, then, 
\vskip-15pt 
\belowdisplayskip=2pt 
\begin{eqnarray}
 t_a(i)&:=& t_b(i) + t_{av}\nonumber \\
 l_a(i)&:=& l_b(i) \nonumber
\end{eqnarray}
where $t_{av}$ is the measure of distance between two nodes when they are within each others' transmission range. Every node performs this update when it comes into contact with another node. Note that $t_a(i)$ is same as temporal distance $t_{ia}$ and is used to approximate $t_{ai}$ at node \textit{a} as described in Section \ref{stdr}. The value of $t_{av}$ is a small constant greater than zero but less than or equal to one time unit (increment by which local timers are updated) i.e. $t_{av}\in(0,1]$ \cite{proximol}. Note that it's a different definition of $t_{av}$ from that in \cite{spyroscp:17} where the value of average time required to travel between the two nodes is included. A small value of $t_{av}$  is used in our experiments as we are interested in the time a request will take in forwarding and not the physical distance between nodes.


\subsubsection{Service advertisement and controlling distribution radius}
Each node keeps track of its temporal distance from all other nodes for the service composition algorithm. To reduce network overhead, temporal distance values of only those nodes are stored that are in close vicinity (e.g.\ nodes with temporal distance less than 15 minutes). The exact threshold value for retaining temporal distance values of other nodes depends on the application requirement in terms of how much delay can be tolerated. This makes our approach scalable for large networks, as nodes have a direct mechanism to filter unnecessary information about devices that are further away. Thus, each time two nodes come into contact, they only update the timers and service load estimates for those remaining nodes in network whose temporal distances (implied proximity of location) are smaller than the threshold value (which is suitable for application requirements). Similar timers can directly be used to control epidemic forwarding of description of services that other nodes provide to a limited space. As network size grows, nodes only maintain a fixed set of values to efficiently select and forward service requests.

\subsection{Levels of awareness}
Note that even though any forwarding scheme can be used to forward service requests efficiently, the selection of service set (and consequently the destination nodes for forwarding) still needs to take into account the likelihood of delivery. Even though our algorithm is independent of any forwarding scheme, selection of a particular service set has a significant impact on performance (service completion rate and delay) of any forwarding algorithm. The selection of a particular service set can be made with different levels of awareness about rest of the network. More specifically, the edge cost between vertices of a service graph is based upon the knowledge of temporal distance and service loads. These levels are summarized in Table \ref{loa} for node \textit{a} where $l_a(i)$ represents estimate for load of \textit{node i} at node \textit{a} and $\widehat{t_i(j)}$ represents estimate for temporal distance $t_i(j)$  at node \textit{a}. Note that, timers maintained by node \textit{a} only provide $t_a(i)$, therefore, different levels of awareness are used to estimate $t_i(j)$ for $i,j\ne a$.  

\begin{table}[!t]
\renewcommand{\arraystretch}{1.2}
\caption{Settings for different levels of awareness}
\label{loa}
\centering
\begin{tabular}{c||c|c|c}
\hline
\bfseries Level & $\widehat{t_a(i)},\widehat{t_i(a)}$   &\bfseries $\widehat{t_i(j)}:i,j\ne a$  & \bfseries $l_a(i)$   \\
\hline\hline
 minimal & 1 & 1 & 0\\
\hline
 local  & $t_a(i)$ & $t_a(i)+t_a(j)$ & $l_a(i)$(delayed)\\
\hline
 global & $t_a(i)$ & $t_i(j)$(delayed) & $l_a(i)$(delayed) \\
\hline
perfect & $t_a(i)$ &$t_i(j)$ &$l_a(i)$\\
\hline
\end{tabular}
\end{table}

At the very basic level it is assumed that a node knows about services being provided in the environment but does not have an estimate of its temporal distances from particular devices. This is reflected as \textit{minimal} in Table \ref{loa} as it requires no housekeeping. In this case, a node randomly selects the device providing the required service(s) for composition.

The next level of \textit{local} knowledge is by use of timers as described in Section \ref{timerx}. A node has knowledge of its temporal distance from other nodes in the network (e.g., $t_a(i)$ for node \textit{a}). A node also knows about the latest service load at other nodes, which is forwarded to it in a distributed manner [Section \ref{timerx}] and therefore incurs some delay. This level requires exchanging $O(N)$ values ($t_a(i),l_a(i)$ $\forall i$ at  node \textit{a}) per contact. In order to update link cost between services provided at other nodes, a node needs to be aware of temporal distances between other nodes (i.e. $t_i(j):i,j\ne a$). However, this information is not available in \textit{local knowledge}. Therefore, a node uses its own timer values for approximation as shown in Table \ref{loa}. For example, node \textit{a} uses $(t_a(i)+t_a(j))$ as an approximation for temporal distance $t_i(j)$ because $|t_a(i)-t_a(j)|<t_i(j)<t_a(i)+t_a(j)$. Even though this approximation (an upper bound on $t_i(j)$) is not accurate, it still gives an idea of proximity of a device. Later we show that the performance results using local knowledge are fairly close to those achieved with perfect knowledge.

The next higher level of \textit{distributed global} knowledge is when a node also receives timers of other nodes  in a distributed manner i.e.\ the estimates of temporal distances between all pairs of nodes which requires exchanging $O(N^2)$ values ($t_i(j),l_a(i)$ $ \forall i,j$ at  node \textit{a}) per contact. However, this information about temporal distance between other nodes is still not up-to-date because of propagation delay. Therefore, we further make comparison with a \textit{perfect} system that assumes centralized knowledge of temporal distances and service loads i.e. all information is available to nodes without any delay.


\section{Performance evaluation}
\label{perfeval}

The key novelty of our proposed solution is to enable service composition over opportunistic paths. To illustrate the effectiveness, we compare the performance with solutions that do not use composition, but wait to encounter a given node providing the entire functionality needed (exact service match). In addition, we also compare with the case where service are composed, but only using other devices within reach through a simultaneous multi-hop path available at the time when service requests are generated. In the former case (exact match) we don't use composition at all. In the latter, we use composition, but we don't use opportunistic networking, i.e. we don't exploit nodes that are temporarily disconnected, but that will become reachable again in the near future. Our results (Section 6.2) show the advantage of jointly compose services, and using opportunistic networking techniques to do so. It is found that the service composition over opportunistic paths completes a higher percentage of service requests and reduces the delay to complete requests. Next we evaluate four versions of our proposed solution and find that maintaining limited local knowledge about the rest of the network is sufficient to achieve very good results (Section 6.3). We also compare using different routing mechanisms (Section 6.5).

To understand the improvement in performance, we look at the breakdown of completed requests (i.e. what percentage of completed services required 1 service, 2 services or more services to complete the request), the distribution of hop counts to see how much further a request travels in the network, and effect of high load, long request timeout durations, and node/service densities (Section 6.4). Finally, in Section 7 we further analyse the performance of service composition, by considering multiple mobility settings (Section 7.1), the advantage of taking into account information about current load of nodes (Section 7.2), the degree of approximation with respect to exact solutions (Section 7.3), the sensitivity to the service component availability (Section 7.4).

We primarily use synthetic traces to control the environment and evaluate in a range of mobility characteristics. Real traces are only used to validate the results from synthetic traces [fig 7] but are otherwise limited as we cannot increase trace duration, node density, or modify other mobility characteristics. Some of those traces have been obtained by logging GPS coordinates of the users, in order to track their mobility behaviour. They are then processed to identify communication opportunities between these mobile users, which is the information we use in our analysis.

\begin{table}[!t]
\renewcommand{\arraystretch}{1.0}
\caption{Simulation parameters}
\label{param}
\centering
\vspace*{-7pt}
\begin{tabular}{c||c}
\hline
\bfseries Description &  Range \textbf{[default value]}   \\
\hline\hline
Number of services 	&	\textbf{[20]}	\\	
\hline
Number of service providers  & \textbf{[20]}, 40 \\
\hline
Repetition of each service 	& 	1, \textbf{[2]}, 3, 4 \\
\hline
Request timeout 	&	10, \textbf{[15]}, 20, 30 \\
\hline
Request rate per node 		&		0.2, \textbf{[0.4]}, 0.67, 1/min \\
\hline
Trace duration			& \bfseries [10 hours] \\
\hline
 Transmission range of device  & \bfseries [100m]\\
\hline
\end{tabular}
\vspace*{-10pt}
\end{table}

\subsection{Simulation setup:} 
\label{simSetup}
The real mobility traces have been collected from participants that carry GPS receivers which log position at 30 second intervals in a State Fair \cite{strace}.  In order to make a comparison with suitable number of nodes, track logs by same user on different days are considered to be a separate user on the same day. These logs have durations from around one to ten hours each day. We truncate all logs to 90 minutes during which 18 users record their location. Synthetic mobility traces are generated using Levy Walk mobility model\cite{levywalk:28} with 20 (10 slow,  10 fast moving) and 40 (30 slow,  10 fast moving) nodes.  The trace settings are same as those defined in \cite{proximol}.

\begin{figure}[!t]
\centering
\includegraphics{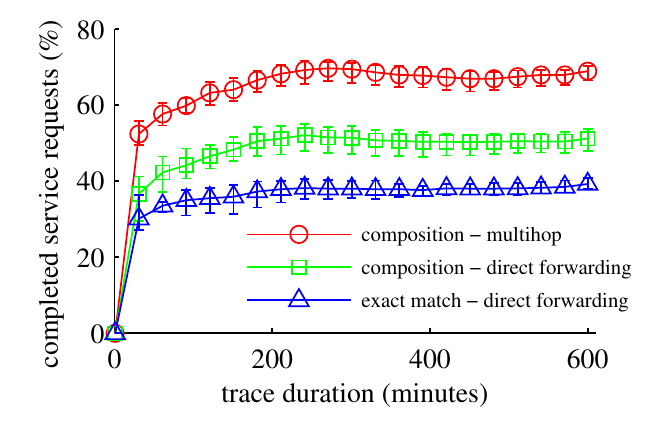}   
\caption{Higher service completion in multi-hop composition}
\label{basicd}
\end{figure}
\begin{figure}[!t]
\centering
\includegraphics{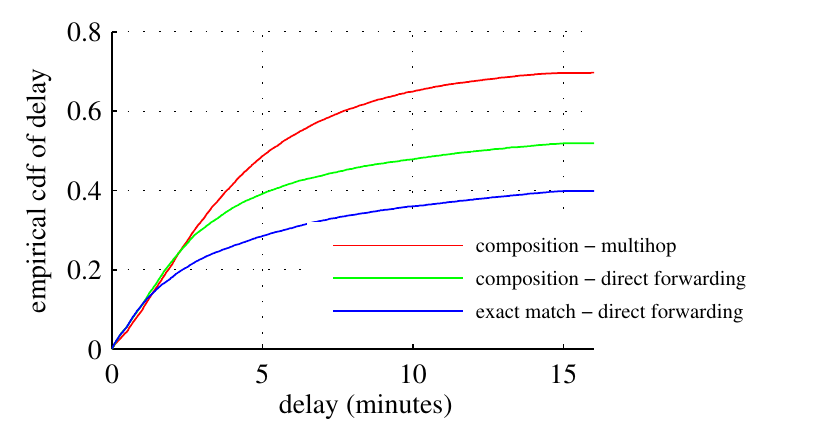}   
\caption{Lower delay in multi-hop composition}
\label{basic2}
\end{figure}

\begin{figure*}[!th]
\centering
\includegraphics{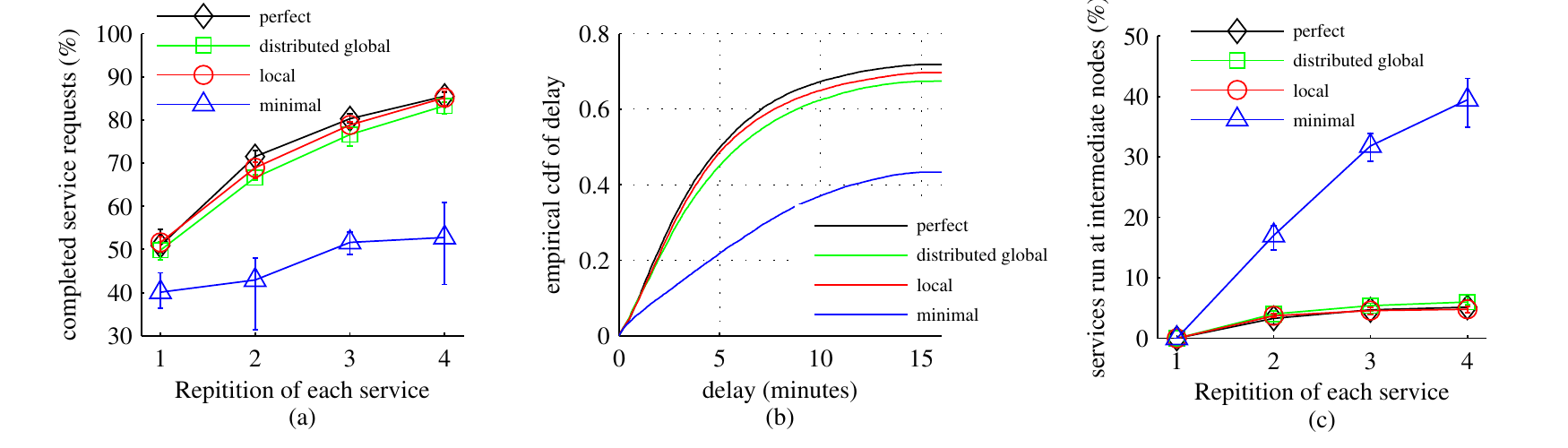}   
\caption{Levy Walk: With different levels of awareness (a) service completion, (b) delay, and (c) services run at intermediate nodes}
\label{awr}
\end{figure*}

All plots show the average of five simulation runs and the error bars are plotted on the minimum and maximum values in those runs. Table \ref{param} summarizes the simulation parameters.  In all following plots, default values of these parameters are used unless it is mentioned otherwise under the plot. Parameters used for forwarding schemes are EBR-Encounter rate window=10 minutes, $t_{av}$=10min for TT and 0.5min for MT. A total of seven input/output types are used to create $C_2^7 =21$ unique services. Service $s_{17}$ is dropped to leave 20 unique services that are provided at 20 nodes in the network. Note that for a service repetition level of 3, all services are provided by 3 unique nodes (randomly selected) such that every node provides exactly 3 services. Service requests are only generated for services with $k\geq 4$ and the average execution time of service is set to 30s. Each service request times out after 15 minutes so that previous service requests do not overload the system. Thus, services are considered incomplete when they are not complete within 15 minutes. In all simulation results, service requests are not generated in the last 15 minutes (duration of request timeout) of the trace duration. We consider a simulation analysis including both a transient and a steady regime phase. The transient regime lasts for approximately the initial 30 minutes, after which all nodes acquire information about network resources and system enters the steady regime. In the transient regime the system starts with no initial service loads. Therefore, delays of only those compositions are considered that are generated once the system is well inside the steady state regime (service requests that are generated after the first two hours of simulation, are considered).

\subsection{Direct and multi-hop forwarding}
Figure \ref{basicd} shows the service completion rate in the Levy Walk mobility trace. In all these cases \textit{local} level of awareness is used to construct a service graph. The search for a single service that exactly matches the request fulfills 40\% of the requests. However, when the condition is relaxed to compose multiple services to meet the requirements, about 50\% of service requests are completed. As a further improvement, when multi-hop forwarding paths are used to forward requests and service outputs, 70\% requests are completed.  This shows significant improvement that service composition with multi-hop forwarding paths achieves compared to other mechanisms \cite{reliable,dependable}. In contrast, our algorithm tolerates disconnections and utilizes paths that become available over time. As a result, in addition to utilizing services at devices with directly connected paths to a node, ones that are present in the nearby vicinity can be invoked at the expense of slightly higher delays. However, in opportunistic networks, multi-hop forwarding   still yields lower delays as compared to direct one-hop forwarding. This is shown in Figure
 \ref{basic2} where 50\% of compositions are made within 5 minutes using multi-hop forwarding, whereas it takes 12 minutes in direct forwarding. In case when only exact match is searched in the network, only 40\% requests are completed after 15 minutes. The CDF plot (proportion of samples less than a value) is based on the aggregate delay per service from five simulation runs. As seen from Figure \ref{basicd}, the percentage of completed services increases sharply in the first 30 minutes (transient regime) and then becomes steady. Therefore, in Figure \ref{basic2} and later, delays of only those compositions are considered that are generated once the system is well inside the steady state regime (after first two hours of simulation).

\subsection{Levels of awareness}
While there is significant improvement in service completion rate and delays when some knowledge about temporal distances from other nodes is used, there is not much difference in the performances of perfect, globally aware and local schemes. This is clear from Figure \ref{awr}a where in a range of service densities, completion rate of perfect, distributed global and local levels of awareness are within 3\% of each other.  Figure \ref{awr}b shows the delays when each service is repeated twice i.e.\ two different nodes provide the same service. The local scheme has delays between the globally aware and perfect scheme as shown in Figure \ref{awr}b.  One explanation for this behavior is that approximation for temporal distance between other nodes is better than delayed information of actual temporal distance between those nodes. Notice that at the time of request initiation, in perfect knowledge, a node knows exact distance between other nodes only at that instant of time -  a node does not know the exact distance between other nodes in future when the first (or second, third etc.) service in composition will be complete. Thus, during the time a forwarded request is received on a device, the network topology changes so much that it does not make much difference if local knowledge was used or perfect knowledge to estimate distances between other nodes for selection of sevices/devices. 
Since, all three schemes have similar performance; we select the scheme with local knowledge for further analysis as it is more light weight (requires exchanging $O(N)$ temporal distance and load values per contact instead of $O(N^2)$). 

In Figure  \ref{awr}c, we validate the selection of correct devices to compose all services. At different service densities, percentage of services that are run at intermediate nodes while they were enroute to some other device  is considered (i.e.\ when required service is opportunistically found at a device different from one selected by the algorithm). While this percentage is quite high around 40\% when minimal knowledge about network is available, it is reduced to under 5\% when some knowledge about network topology is used. This validates that our service selection algorithm selects the correct devices to execute services in a composition. 

\begin{figure}[!t]
\centering
\includegraphics{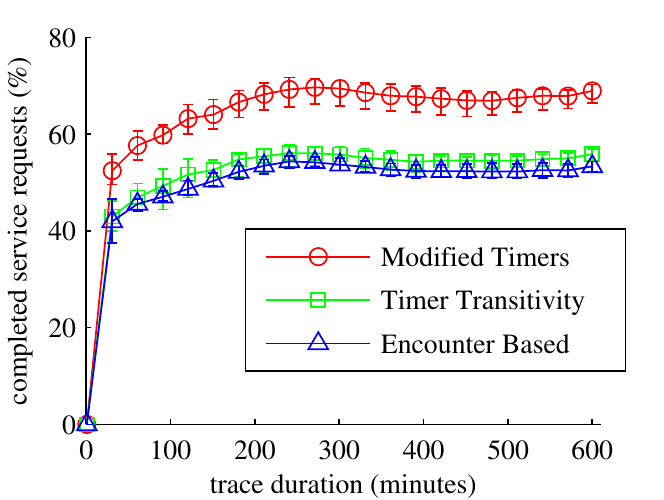}   
\caption{MT  with better service completion}
\label{forw}
\end{figure}

\begin{figure*}[!ht]
\centering
\includegraphics{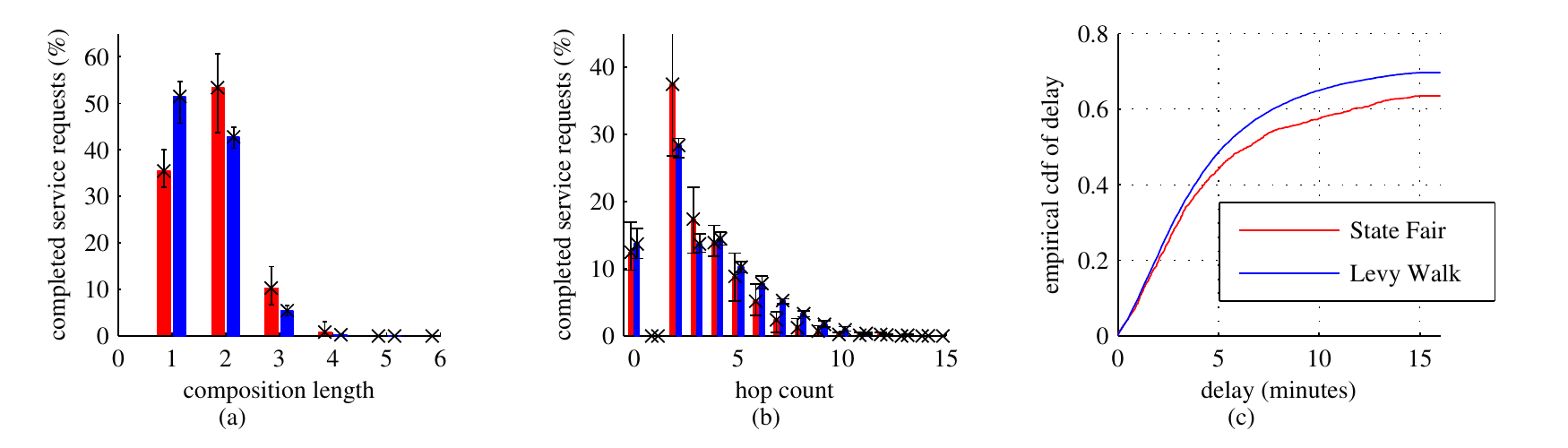}   
\caption{State Fair  [red] and Levy Walk [blue] mobility trace: (a) composition length, (b) hop count, and (c) delay profiles for multi-hop service composition}
\label{profilelw}
\end{figure*}

\begin{figure*}[!t]
\centering
\includegraphics{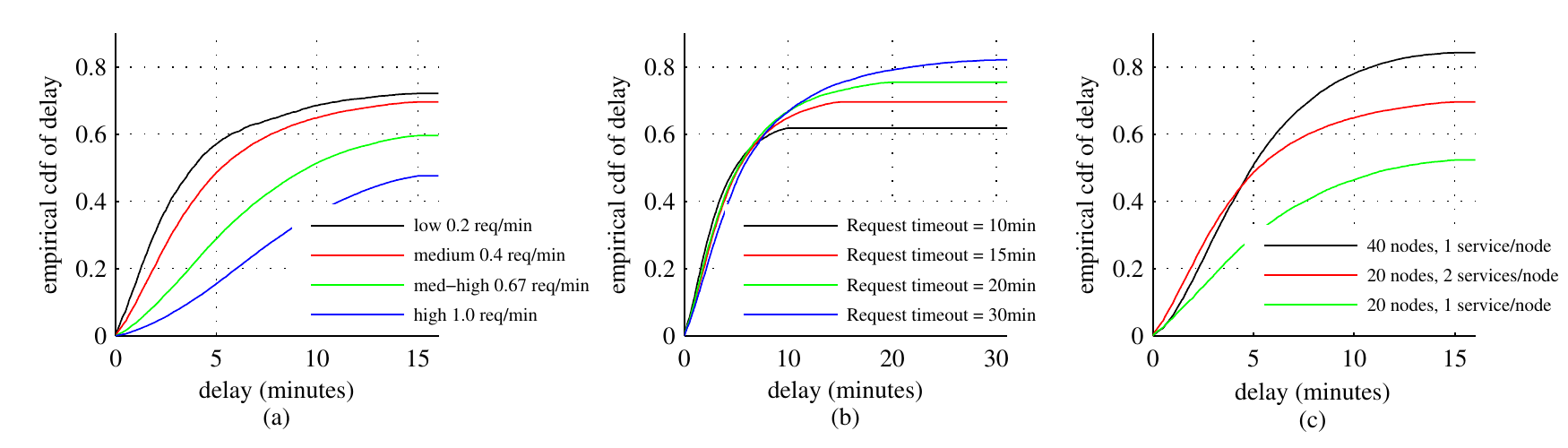}   
\caption{Levy Walk: Higher service completion and lower delay at (a) low loads, (b) long request timeout, and (c) high node, service density for multi-hop service composition}
\label{figallb}
\end{figure*}

\subsection{Composition length, delay, and hop profiles}
Figure \ref{profilelw} shows the detailed profile of multi-hop service composition for State Fair and Levy Walk mobility trace. Since, State Fair trace lasts only 90 minutes, we have performed extensive simulation on Levy walk with a duration of 10 hours to evaluate performance results in steady state. The results in this section are used to demonstrate that valuations based on Levy Walk traces are close to real mobility traces in environments similar to state fair. 

Most compositions are comprised of 1 or 2 services and then there are some of three, and four services as shown in Figure \ref{profilelw}a. In our analysis, services are distributed such that every service is provided by at least one node in the network. However, as described earlier, restricting the completion to the exact match degrades performance. Therefore, all compositions (around 60\%) of length more than 1  that are shown in Figure \ref{profilelw}a are optional, and are made to achieve higher service completion rates and minimal delays. However, these compositions come at the expense of a few extra hops as shown in  Figure \ref{profilelw}b. Most of the times (about 80\%) complete services are composed by using less than 5 hops. Around 15\% of requested services are found at the device itself. This causes a hop count of zero as shown in Figure \ref{profilelw}b. If the service is not found at the requesting devices, service completion incurs at least 2 hops (one to send request and one to receive results). Figure \ref{profilelw}c shows the empirical CDF of the time taken from service request to receiving composed results, routed back from the last node in the composition path. Around 60\% of completed service requests have a delay of less than 10 minutes whereas close to 50\% of these are completed in less than 5 minutes. Each service is repeated twice in this run. For higher number of repetitions, delays are lower and percentage of completed services reaches close to 100\% (85\% for four repetitions, see Figure \ref{awr}a). 
In effect, the above results demonstrate that significant service composition results can be obtained in opportunistic environments with reasonable service densities. 

\subsection{Effect of forwarding scheme, request load, request timeout, node and service density}
Figures \ref{forw}  shows the service completion rate of 3 different underlying forwarding schemes used in the service selection algorithm. Only single copy of each request is forwarded. MT shows better delivery rate (about 30-40\% higher than other schemes).  Network performance is also analyzed under different service load conditions and levels of connectivity. Requests rates are varied from 0.2 to 1 request per minute per node. Figure \ref{figallb}a shows that service completion rate decreases under high loads. At high node/service density Figure \ref{figallb}c shows that almost 85\% of all requests can be composed. Also, more requests can be completed when a service request does not timeout for a longer time [Figure \ref{figallb}b]. Thus, while service compositions can be made in a moderately connected network, our algorithm adapts well in sparse scenarios with higher service request loads and short request timeout duration. Note that variation in load greatly changes the delay profile under 5 min as shown in Figure \ref{figallb}a. This is because higher loads result in longer service queues and services take longer to execute even when a connected path to service provider exists. Whereas variation in request timeout changes the delay profile at over 10 min as shows in Figure \ref{figallb}b. This is because longer request timeouts keep the service requests in the queue which are otherwise dropped without completion. As a side effect, average service load at nodes increases slightly. For this reason, 40\% of compositions are made in 4.5 min with request timeout of 30min but only in 3.5min with timeout of 10 min due to change in average load at nodes. Figure \ref{figallb}c shows that average delays are lower when 40 services are distributed in 40 nodes instead of 20 nodes even though the number of generated requests is double in a network with 40 nodes. This is because the network with 40 nodes is more dense and connected. 

\begin{figure*}[!t]
\centering
\includegraphics{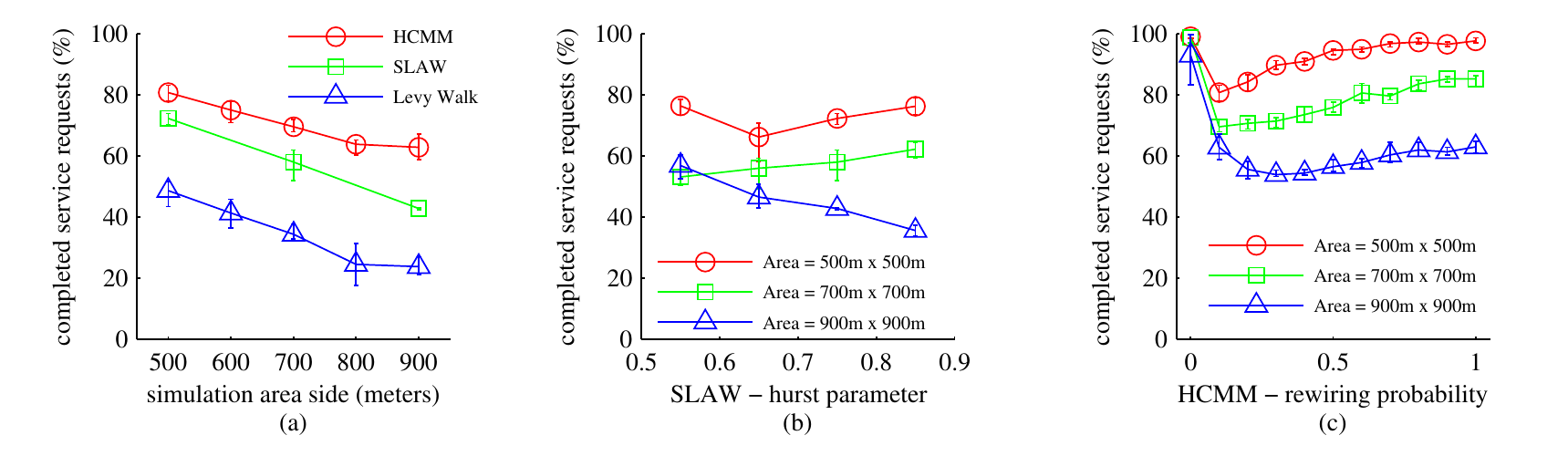}   
\caption{Performance of service composition under varying mobility parameters in  (a) Levy Walk, (b) SLAW, and (c) HCMM mobililty models.}
\label{2_mobParameters}
\end{figure*}

\section{Characteristics of the algorithm}
\label{xcharcal}

While the previous section shows that service composition results in improved performance when opportunistic paths are used, it does not show the limitations or the key underlying features that result in such improved performance and what can be done further to achieve an even better performance. Thus we analyze performance in different mobility environments - communities, clusters and uniformly distributed. Within each environment, we modify the underlying parameters that impact mobility and discuss their impact. To understand the variation in performance  under different mobility environments, we examine the difference between the composition cost computed in the service graph and the actual cost of completing a service.  This difference would be an indication of the accuracy of the temporal distance as a measure of device reachability. We further investigate that load aware composition is useful only in more connected/dense networks. To improve the performance even further, it is useful to replace less popular services with more popular service when the service request patterns are non-uniform (i.e. some services are requested a lot more often that others). 

To compare performance under different mobility characteristics, we consider all nodes to be moving at the same speed (unlike the slow and fast moving nodes) but vary the key distinctive feature of three different mobility models. Similarly, to analyze the effect of composition length (number of services to be composed to complete a request), we run simulations for requests  made {\em only} for services of compositions of length one and compare against a separate simulation run for requests made {\em only} for services of composition length two and so on. 

\subsection{Mobility characteristics}
\label{mobCharac}
Fig.~\ref{2_mobParameters} shows performance of service composition under varying mobility parameters in Levy walk, SLAW and HCMM mobility models. 
These models provide a wide range of mobility characteristics to thoroughly evaluate performance of service composition: HCMM - movement in communities; SLAW - a clustered environment; and Levy walk - uniformly distributed mobile users. The key distinctive feature of SLAW is the Hurst Parameter that controls the degree of self-similarity and clustering in the environment. The flight length and pause time distribution are similar in Levy walk and SLAW so we do not vary those parameters. The key distinctive feature of HCMM is the rewiring probability which establishes the frequency of back and forth travel of nodes between communities. For all these mobility models we also study the effect of node density by keeping the number of nodes constant but varying the simulation area. 

As illustrated in Fig.~\ref{2_mobParameters}a, with similar user densities, more services are completed in HCMM as compared to SLAW and Levy Walk.
For example, with 20 users in a simulated area of 700m$\times$700m, 70\% services are completed in HCMM as compared to 58\% in SLAW and 35\% in Levy Walk. This is because nodes stay closest to each other in communities under HCMM, are still closer in clusters under SLAW as compared to uniform distribution in the whole simulation area under Levy walk. 

Fig.~\ref{2_mobParameters}a also shows that percentage of completed services decreases with user density; e.g.\ completed services in HCMM decreases from 81\% to 64\% when simulation area increases from 500m$\times$500m to 900m$\times$900m. A similar trend is followed for SLAW and Levy walk.

Fig.~\ref{2_mobParameters}b shows the effect of Hurst Parameter. There is no significant trend at high node densities but the performance of service composition decreases at lower user density (20 users in 900m$\times$900m area). From visual inspection of trajectory plots, Hurst parameter of 0.85 shows longer flights spread in larger area than Hurst parameter of 0.55. However, in several random generations of mobility plots we have found a great variation in distribution of waypoints for the same value of Hurst parameter. Also, since most real world traces have Hurst parameter value close to 0.75 \cite{slawg,slaw:29} we use this value for all the traces when performance under SLAW is compared with Levy walk or HCMM.

Fig.~\ref{2_mobParameters}c shows the effect of rewiring probability in HCMM. Initially there is a sharp decrease and then an increase in completed services when rewiring probability is increased from zero. This is because, more services are completed when all nodes stay in the same community (rewiring probability $=$ 0) versus when some of the nodes leave before completing the pending requests (rewiring probability $=$ 0.1). However, as the back and forth travel of nodes between communities increases (rewiring probability $>$ 0.4), there is an additional advantage of having services from a different community to be available. As a result, with increase value of rewiring probability, there is again an improvement in percentage of completed services. However, the advantage of travel between communities also depends on how far the communities are. Thus, least number of services are completed at rewiring probability of 0.1 in 500m$\times$500m area but in a larger area of  900m$\times$900m it occurs at 0.4. The advantage of traveling kicks in at higher rewiring probability (0.4 instead of 0.1) in 900m$\times$900m because the communities are further apart and it takes more time to travel between communities. Thus, a small amount of travel to other communities decreases performance, but a lot of travelling between communities again improves performance of service composition. In all simulations, we use the value of rewiring probability to be 0.1 when comparing performance against Levy walk and SLAW.

\subsection{Load estimation}
\label{load_est}
\begin{figure}[!t]
\centering
\includegraphics[width=2.5in]{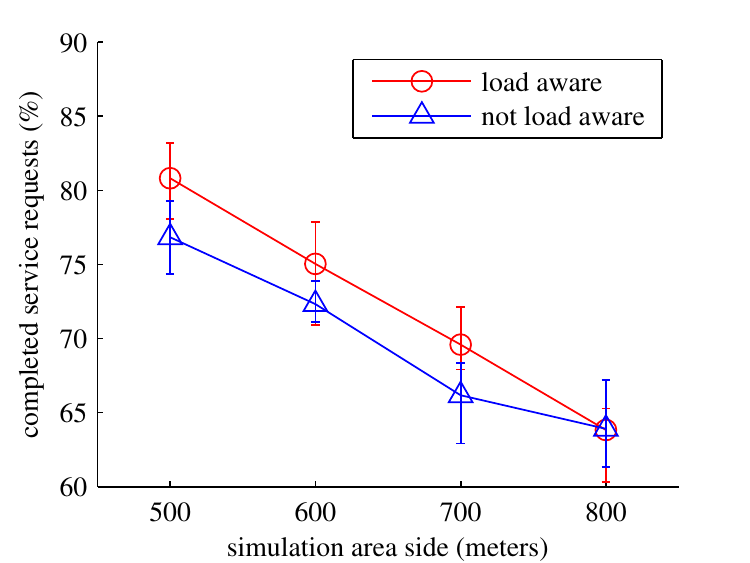}   
\caption{ Load--aware service composition (LA) improves performance in a more connected network, but does not make much difference in sparsely connected networks}
\label{5_loadAware}
\end{figure}

We explore performance of service composition algorithm with and without an estimate of load i.e. we consider 
\begin{IEEEenumerate}
\item LA: load aware service composition (algorithm that takes into account the temporal distance as well as load at other nodes), and
\item NLA: not load aware service composition (algorithm that takes into account the temporal distance and but does not take into account the load at other nodes)
\end{IEEEenumerate}
Fig. \ref{5_loadAware} shows that load aware service composition (LA) improves performance by 10\% in a more connected network (20 users in small area 500m$\times$500m), but does not make much difference in sparsely connected networks (20 users in large area 900m$\times$900m) in HCMM mobility. LA performs better than NLA in more connected networks, because delay due to sparse contacts is comparable to delay due to load at individual nodes. LA performs similar to NLA in sparse networks because delay due to sparse contacts is more dominant than delay due to load at individual nodes, and knowledge about load does add much value in selecting which services to compose. Thus, the requirement to keep track of load at other nodes is critical when the forwarding delays are comparable to the delay due to queuing and execution of service requests.


\begin{figure}[!t]
\centering
\includegraphics[width=2.5in]{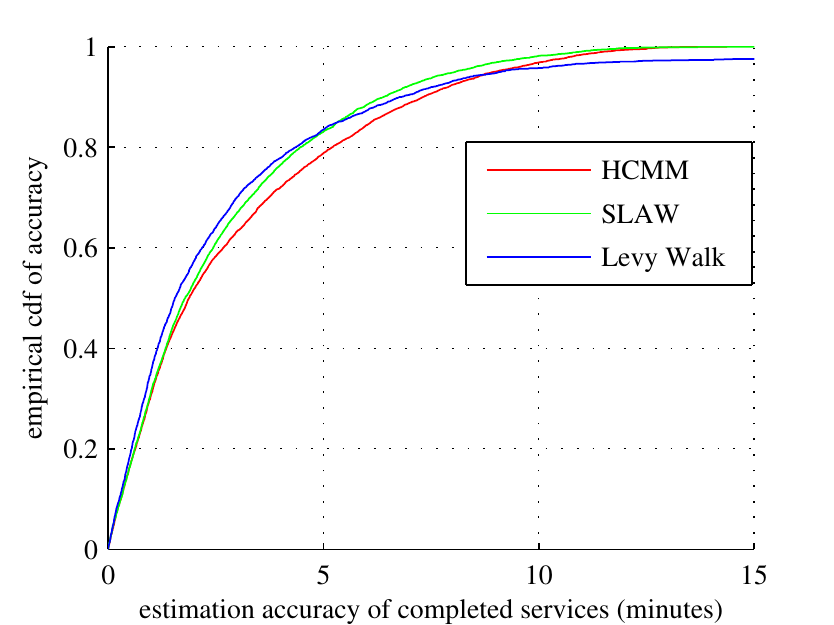}   
\caption{ Estimated cost for 70--80\% of completed services is within 4 minutes of actual delay  in HCMM, SLAW and Levy walk mobility models.}
\label{6_costEstAn}
\end{figure}

\subsection{Estimated cost and actual delay}
\label{perfGuarant}

In this section, we analyze the accuracy of the values used to estimate the time required to complete a service as described in Section \ref{sercomp}. Figure~\ref{6_costEstAn} shows that estimated cost for 70--80\% of completed services is within 4 minutes of actual delay in HCMM, SLAW and Levy walk mobility models. Also, the difference between estimated cost and actual delay is less than 2 minutes for 40\% of the completed services in all three models. This shows that the distributed \textit{local} knowledge algorithm for service composition is fairly accurate in estimation and selection of optimal services that can be composed to complete the request.

\begin{figure}[!t]
\centering
\includegraphics[width=2.5in]{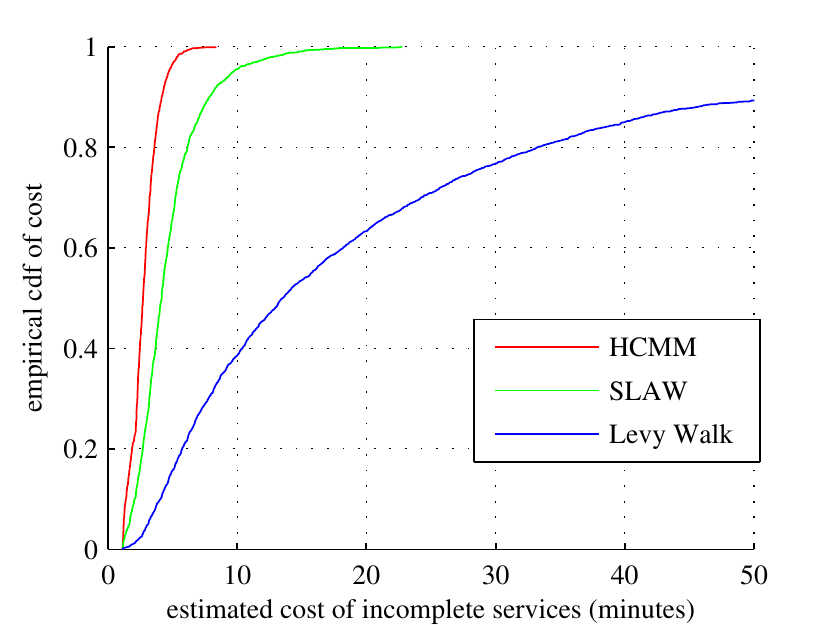}   
\caption{ Estimated cost of incomplete services is 50\% accurate (cost \textgreater 15 minutes) in Levy Walk but not accurate in HCMM and SLAW mobility models.}
\label{6_costEstBn}
\end{figure}

Fig.~\ref{6_costEstBn} shows that estimated cost of incomplete services is 50\% accurate (i.e.\ cost is greater than 15 minutes for 50\% of incomplete services\footnote{Note that services are considered incomplete when they are not complete within 15 minutes. Thus, cost estimate of more than 15 minute is considered accurate for incomplete services}) in Levy Walk but not accurate in HCMM and SLAW mobility models. Notice, that this is not the overall accuracy, but the accuracy of {\em only} incomplete service requests which are a small percentage of the total number of service requests.

Still, this demonstrates that temporal distance is not a very good measure when the user movements contain structures of clusters and communities but has reasonable accuracy when user movement is more evenly spread in the surrounding area. The reason for inaccuracy in HCMM and SLAW is that nodes leave the community at random which results in large number of incomplete services. Using temporal distance alone does not predict which nodes may or may not leave the community/cluster in future under HCMM/SLAW. The possible solutions in such a case is to either introduce redundancy by requesting services with more than one nodes or to have a mechanism to predict user movement (i.e.\ which users may or may not leave the community) through other sources of information (user schedule, history of visits etc.).

\begin{figure*}[!t]
\centering
\includegraphics{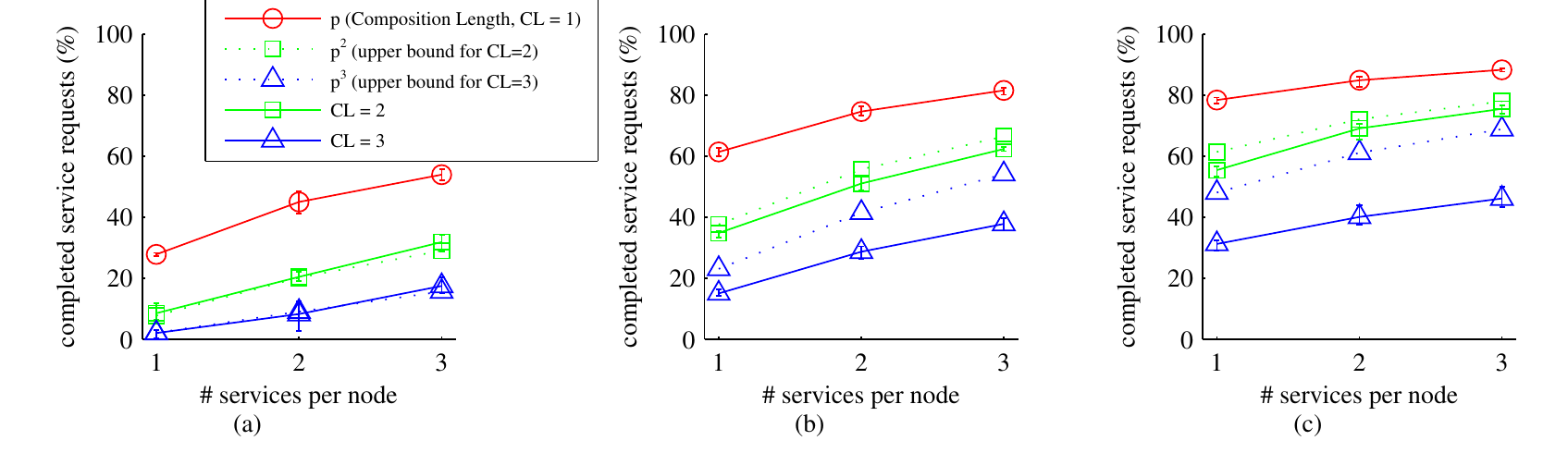}   
\caption{Sensitivity analysis for composition of 1,2 or 3 services in: (a) Levy Walk, (b) SLAW, and (c) HCMM mobility models.}
\label{3_sensitivity}
\end{figure*}

\subsection{Sensitivity to composition length and service distribution}
\label{sensitivity}
The required number of services that are composed to complete a request greatly affects the performance of service composition. In addition, performance improves when services are distributed in the same pattern as the requests are generated (i.e. more commonly requested services are provided at more nodes than the rarely requested services). To analyze only the effect of composition length and service distribution, a slightly different set of services is used from the one described in Section \ref{simSetup}. A total of 20 input/output types are used to create 20 unique services of functionality equal to one $k=1$. Thus, the created services are $\{s_{1\,2},s_{2\,3},s_{3\,4},...,s_{8\,9}, s_{9\,10}, s_{10\,11},..., s_{19\,20},  s_{20\,1} \}$. Each of these services is provided by two nodes selected uniformly at random.

\subsubsection{Composition length}
\label{senCL}
Fig. \ref{3_sensitivity} shows sensitivity analysis for composition of 1,2 or 3 services in Levy walk, SLAW and HCMM mobility models for a range of service densities. The performance of service composition is analyzed with different request patterns as follows: Service requests require
\begin{IEEEenumerate}
\item exactly a single service -- composition length = 1, 
\item composition of two services -- composition length = 2, 
\item composition of three services -- composition length = 3. 
\end{IEEEenumerate}
For example, a request with input 3 and output 5 requires two services $s_{3\,4}$ and $s_{4\,5}$. The simulation results show that the performance degrades when more services are required to complete a request. It should be noted, that this result actually elaborates the earlier deduction in Figure~\ref{basicd} where composition leads to better performance.
The difference between these two results is in the simulation setting. Figure Figure~\ref{basicd} shows the results for a general service distribution and request pattern [Section \ref{simSetup}] - in this setup, allowing for composition of more than one service significantly improves the percentage of completed services. 

In contrast, Figure~\ref{3_sensitivity} shows a complete simulation run where services with a fixed composition length are requested. Thus, it shows the percentage of completed services by making requests that only require a single service. Figure~\ref{3_sensitivity} then shows the impact of composition by making requests that require composition of two services in a separate simulation run. More specifically, the percentage of completed services decreases when two or more services are composed {\em for every request}. This is to emphasize the point that composition by itself does {\em not} lead to better performance [Figure~\ref{3_sensitivity}] but rather composition leads to better performance only when the cost of selected composition is compared and found better against the cost of all other possible compositions [Figure~\ref{basicd}] like in our service composition algorithm. 

Figure~\ref{3_sensitivity} shows sensitivity to composition length in different mobility environments with a controlled pattern of service requests. Specifically, if the ratio of completed services is $p$ ( where $0\leq p \leq 1$) when composition length = 1, then the ratio of completed services is less than $p^2$ when composition length = 2, and is less than $p^3$ when composition length =3. This can be explained by assuming the probability of completing a single service to be $p$. Assume that probability to complete a service is independent of probability to complete other services in the request, the particular service that is requested and the node that requests it. Then probability of completing a request for composition two services is $p^2$ and probability of completing a request for composition two services is $p^3$. 
In simulation, the actual ratio of complete services is slightly under $p^2$ and $p^3$ because available time--per--service reduces when a request needs more than one services to be composed. Thus, probability of completing a request for composition of two services is slightly less than $p^2$ due to smaller available time--per--service. 

The value of $p^2$ and $p^3$ serves as an upper bound (estimate) to the ratio of completed services for composition length = 2 and 3 respectively. This upper bound is close (slightly greater than actual ratio of completed services) under Levy Walk Fig.~\ref{3_sensitivity}a, but is not close (a lot greater than actual ratio of completed services) in SLAW [Fig.~\ref{3_sensitivity}b] and HCMM [Fig.~\ref{3_sensitivity}c]. This is because nodes and services are located in clusters/communities in SLAW/HCMM. Unlike levy walk, the probability of completing a service is greatly affected by whether the requested service is found in the same cluster or community in SLAW/HCMM. When a single service is requested, most of the completed services are located within the same cluster/community in SLAW/HCMM. Therefore, the estimate of completing a single service is biased towards assumption of availability of service in the same or nearby cluster/community. As a result, whenever one or more services required to complete the request are not available in the same cluster or community, the probability of completing the request becomes way lower than the upper bound of $p^2$ and $p^3$ for the composition length of 2 and 3 respectively.

\subsubsection{Service distribution}
\label{servDist}
\begin{figure}[!t]
\centering
\includegraphics[width=2.5in]{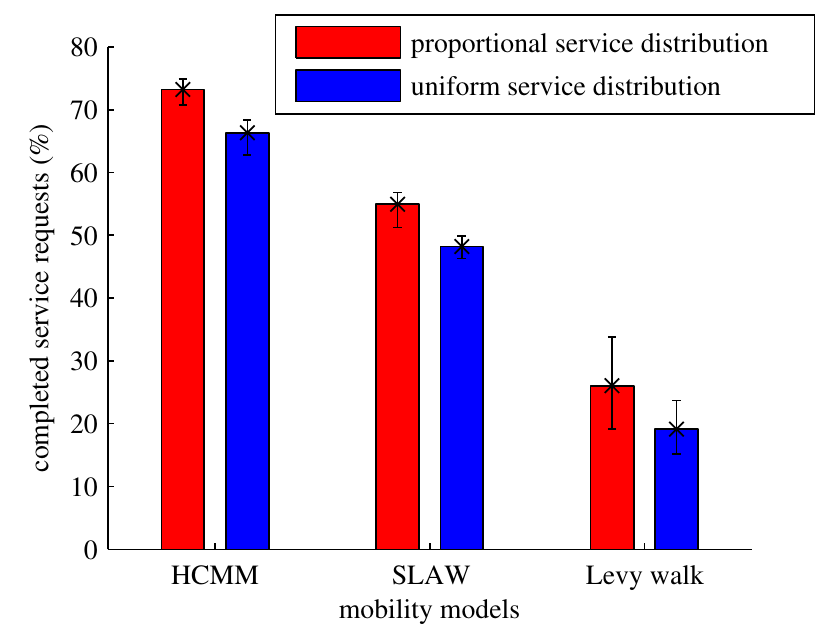}   
\caption{Performance of service composition improves if less popular services are replaced by more   popular services.}
\label{4_serviceDist}
\end{figure}

Figure~\ref{4_serviceDist} shows that performance of service composition improves if less popular (requested) services are replaced by more   popular (requested) services. To analyze sensitivity to service distribution consider that half of the services (10 out of 20 services) are requested $3$ times as often as the other half. Then following two scenarios are of interest:
\begin{IEEEitemize}
\item Uniform distribution: Each service (of all 20 services) is provided at two different nodes. The total number of services including repetition is 40 such that every node provides two services.
\item Proportional distribution: Each of the 10 services (that are requested more often) is provided at three different nodes, and each of the 10 services (that are requested less often) are provided at only one node. The total number of services including repetition is still 40 such that every node provides two services.
\end{IEEEitemize}

Also, services and nodes are selected uniformly at random for each selection. Requests are made such that they require a composition of two services.  Results show an improvement of close to 20\% when services are distributed proportional to request rate as demonstrated in Figure~\ref{4_serviceDist} under Levy walk, SLAW and HCMM. This provides a key insight that replacement of less popular services with more popular services to match the request pattern increases percentage of completed services. The node on which service is replaced is randomly selected out of the two nodes providing the less popular service. Performance can be improved even further by considering location and mobility characteristics of a node on which service is replaced.

\section{Conclusion and future work}
\label{xconclusion}
In this paper, we proposed a novel algorithm for service composition in opportunistic networks to provide mobile users  the ability to benefit from a larger set of services available in the local environment. The proposed algorithm makes efficient service selections from devices that are in close proximity. Key insight used in the solution is that temporal distance between devices provides a direct measure for reachability of nodes when an end-to-end connected path does not exist. This way, based on a service graph, a composition sequence can be selected to meet the requirements of service request while any routing scheme can be used to forward service parameters. 

Through extensive simulations on real and synthetic mobility traces, we conclude that using only local knowledge about temporal distances and service loads, a large percentage of services can be composed in an opportunistic network using multi-hop paths among devices. In order to complete a request, searching for possible compositions is better than looking for a single service that matches the request exactly. Performance of service composition  improves (more requests are completed with lower delays)  under multihop paths, high service density, high node density, clustered environments with same node density (like HCMM or SLAW in comparison with Levy walk), long request timeout duration  and low request rate. In scenarios when some services are requested more often than others, performance also improves when less popular (requested) services are replaced by more popular (requested) services.  

Future work will explore key insights about replacement of less popular services with more popular services - specifically based on the location and mobility characteristics of the user providing a particular service. Additionally, replication of service requests can be explored to reduce delays and improve percentage of completed service. Another direction of work can be to explore the effect of dropping service requests when expected delay is higher than the application requirements to offload the network resources.




\section{Acknowledgments}
The research work presented in this paper was part of Umair Sadiq's Doctoral Work at the University of Texas at Arlington. The research work was carried out with support from the US National Foundations Grants ECCS-0824120, CSR 0834493 and European Commission under the FP7 MOTO (FP7-317959), FP7 EINS (FP7-288021) and EIT ICT Labs MOSES (Business Plan 2014) projects. The authors would like to thank Chiara Boldrini for help with HCMM model and Sidra Bashir for help in running simulations.
%

\bibliographystyle{IEEEtran}



%
%





\begin{IEEEbiography}[{\includegraphics[width=1in,height=1.25in,clip,keepaspectratio]{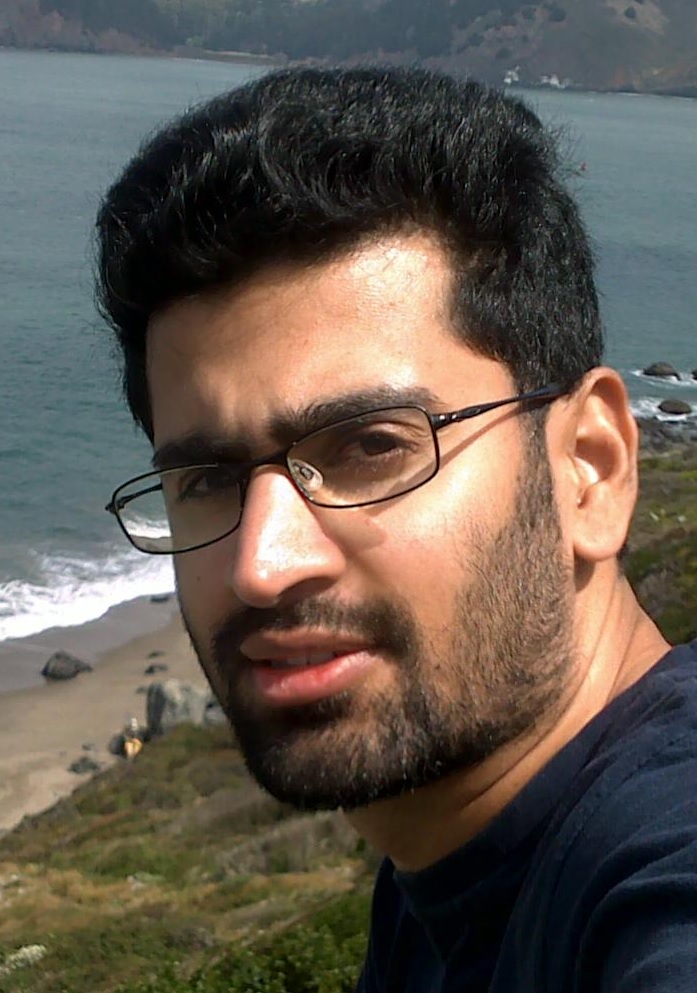}}]{Umair Sadiq}
received the BS degree in electronics engineering from Ghulam Ishaq Khan Institute, Pakistan and the PhD degree in computer engineering from The University of Texas at Arlington in 2007 and 2013 respectively.

He is currently a software engineer at Sabre. His key interests and expertise include mobile networks, machine learning, data analytics, and business strategy.

Dr.~Sadiq received the John Steven Schuchman Award for outstanding research by a doctoral student at University of Texas at Arlington in 2012.  He received the second best student paper award at IEEE MASS 2011 and was nominated for the best paper award at ACM MSWiM 2011.  He was the recipient of the Ghulam Ishaq Khan Gold Medal for Best Academic Performance in 2007.  He has received Pakistan's national awards for best performance in National Physics Talent Contest in 2003 and for best engineering graduate in 2007. He received special prize for best performance in newly participating countries and a Bronze Medal at the International Physics Olympiad in 2003.
\end{IEEEbiography}

\begin{IEEEbiography}[{\includegraphics[width=1in,height=1.25in,clip,keepaspectratio]{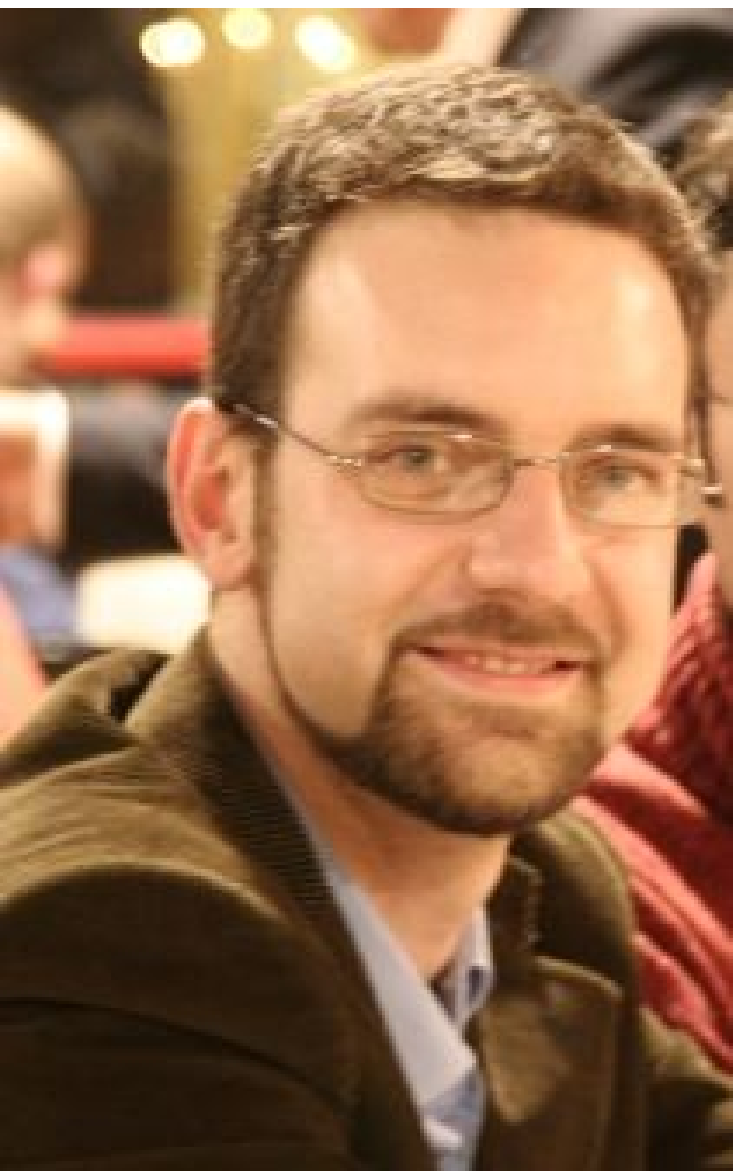}}]{Andrea Passarella}
(PhD in Comp. Eng. 05) is with IIT-CNR, Italy. He was 
a Research Associate at the Computer Laboratory, Cambridge, UK. He 
published 100+ papers on mobile social networks, opportunistic, ad hoc 
and sensor networks, receiving the best paper award at IFIP Networking 
2011 and IEEE WoWMoM 2013. He was PC Co-Chair of IEEE WoWMoM 2011, 
Workshops Co-Chair of IEEE PerCom and WoWMom 2010, and Co-Chair of 
several IEEE and ACM workshops. He is in the Editorial Board of Elsevier 
Pervasive and Mobile Computing and Inderscience IJAACS. He was Guest 
Co-Editor of several special sections in ACM and Elsevier Journals. He 
is the Vice-Chair of the IFIP WG 6.3 Performance of Communication 
Systems.
\end{IEEEbiography}
\vfill

\begin{IEEEbiography}[{\includegraphics[width=1in,height=1.25in,clip,keepaspectratio]{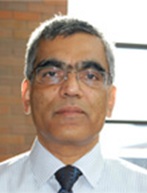}}]{Mohan Kumar}
received the BE (1982) degree from Bangalore
University in India and the MTech (1985) and PhD (1992) 
degrees from the Indian Institute of Science. He is a Professor 
and the Chair of the Department of Computer Science at the 
Rochester Institute of Technology. Previously, he held faculty 
positions at the University of Texas at Arlington and the Curtin 
University, Perth, Australia. His current research interests are 
in pervasive and mobile computing, opportunistic networking 
and computing, sensor networks and distributed computing. He 
has published more than 170 articles in refereed journals and 
conference proceedings and supervised several doctoral 
dissertations and masters theses in the above areas. He has 
developed or co-developed algorithms/architectures for service composition in pervasive 
environments, information acquisition, dissemination and fusion in pervasive and sensor 
systems, and caching and prefetching in mobile, distributed, pervasive, and P2P systems. 
He is one of the founding editors of the Elseviers Pervasive and Mobile computing 
Journal and is one of the area editors of Computer Communications. He has co-guest 
edited special issues of several leading international journals. He is a cofounder of the 
IEEE International Conference on Pervasive Computing and Communications (PerCom), 
where he served as program chair (2003) and general chair (2005). He has also served on 
the technical program committees of numerous international conferences/workshops. He 
is a senior member of the IEEE.
\end{IEEEbiography}

\begin{IEEEbiography}[{\includegraphics[width=1in,height=1.25in,clip,keepaspectratio]{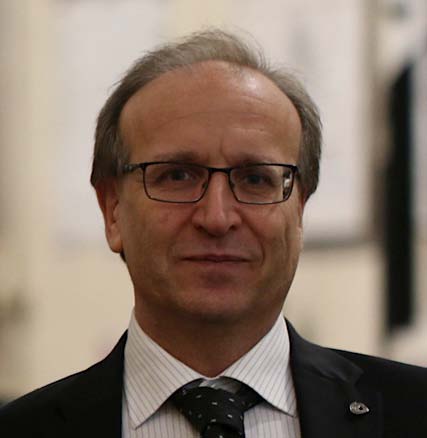}}]{Marco Conti}
is a Research Director of the Italian National Research Council (CNR). He has 
published in journals and conference proceedings more than 300 research papers related to design, 
modelling, and performance evaluation of computer networks, pervasive systems and social 
networks. He co-authored the book Metropolitan Area Networks (MANs): Architectures, Protocols 
and Performance Evaluation (Springer 1997), and he is co-editor of the books: Mobile Ad hoc 
networking: the cutting edge technologies, (IEEE-Wiley 2013), "Mobile Ad Hoc Networking" 
(IEEE-Wiley 2004), and Mobile Ad Hoc Networks: from Theory to Reality (Nova Science 
Publishers 2007). He is Editor-in-Chief of Elsevier Computer Communications journal and 
Associate Editor-in-Chief of Elsevier Pervasive and Mobile Computing journal. He received the 
Best Paper Award at IFIP TC6 Networking 2011, IEEE ISCC 2012 and IEEE WoWMoM 2013. 
He served as TPC chair for several major conferences -- including IFIP Networking 2002, IEEE 
WoWMoM 2005, IEEE PerCom 2006, and ACM MobiHoc 2006 -- and he was general chair 
(among many others) for IEEE WoWMoM 2006, IEEE MASS 2007 and IEEE PerCom 2010. He is the founder of successful conference and workshop series, such as ACM RealMAN, IEEE AOC, ACM MobiOpp, and IFIP/IEEE SustainIT.
\end{IEEEbiography}
\vfill
\end{document}